\documentclass[sigconf]{acmart}

\usepackage[capitalize,noabbrev]{cleveref}
\usepackage{color}
\usepackage{multirow}
\usepackage{subcaption}

\AtBeginDocument{%
  }

\copyrightyear{2024} 
\acmYear{2024} 
\setcopyright{acmlicensed}\acmConference[IUI '24]{29th International Conference on Intelligent User Interfaces}{March 18--21, 2024}{Greenville, SC, USA}
\acmBooktitle{29th International Conference on Intelligent User Interfaces (IUI '24), March 18--21, 2024, Greenville, SC, USA}
\acmDOI{10.1145/3640543.3645206}
\acmISBN{979-8-4007-0508-3/24/03}

\begin{document}

\title{Accuracy-Time Tradeoffs in AI-Assisted Decision Making under Time Pressure}

\author{Siddharth Swaroop}
\orcid{0009-0009-5345-8844}
\affiliation{%
  \institution{Harvard University}
  \city{Boston}
  \state{Massachusetts}
  \country{USA}}
\email{siddharth@seas.harvard.edu}

\author{Zana Bu\c cinca}
\orcid{0000-0002-2644-6065}
\affiliation{%
  \institution{Harvard University}
  \city{Boston}
  \state{Massachusetts}
  \country{USA}}
\email{zbucinca@seas.harvard.edu}

\author{Krzysztof Z. Gajos}
\orcid{0000-0002-1897-9048}
\affiliation{%
  \institution{Harvard University}
  \city{Boston}
  \state{Massachusetts}
  \country{USA}}
\email{kgajos@eecs.harvard.edu}

\author{Finale Doshi-Velez}
\orcid{0000-0003-2886-3898}
\affiliation{%
  \institution{Harvard University}
  \city{Boston}
  \state{Massachusetts}
  \country{USA}}
\email{finale@seas.harvard.edu}

\renewcommand{\shortauthors}{Swaroop et al.}

\begin{abstract}
  In settings where users both need high accuracy and are time-pressured, such as doctors working in emergency rooms, we want to provide AI assistance that both increases decision accuracy and reduces decision-making time. 
Current literature focusses on how users interact with AI assistance when there is no time pressure, finding that different AI assistances have different benefits: some can reduce time taken while increasing overreliance on AI, while others do the opposite. 
The precise benefit can depend on both the user and task. 
In time-pressured scenarios, adapting when we show AI assistance is especially important: relying on the AI assistance can save time, and can therefore be beneficial when the AI is likely to be right. 
We would ideally adapt what AI assistance we show depending on various properties (of the task and of the user) in order to best trade off accuracy and time. 
We introduce a study where users have to answer a series of logic puzzles. 
We find that time pressure affects how users use different AI assistances, making some assistances more beneficial than others when compared to no-time-pressure settings. 
We also find that a user's overreliance rate is a key predictor of their behaviour: overreliers and not-overreliers use different AI assistance types differently. 
We find marginal correlations between a user's overreliance rate (which is related to the user's trust in AI recommendations) and their personality traits (Big Five Personality traits). 
Overall, our work suggests that AI assistances have different accuracy-time tradeoffs when people are under time pressure compared to no time pressure, and we explore how we might adapt AI assistances in this setting.

\end{abstract}



\received{20 February 2007}
\received[revised]{12 March 2009}
\received[accepted]{5 June 2009}

\maketitle

\section{Introduction}
\label{sec:introduction}

Artificially intelligent (AI) systems are being used to help people make decisions in many settings, ranging from helping doctors in disease diagnosis (\citep[e.g.,][]{musen2014clinical}) to helping judges make pretrial-release decisions (\citep[e.g.,][]{green2019principles}).  However, it is increasingly well-understood that different situations and different people may require different forms of AI assistance~\citep{arshad2015investigating, fogliato2022goes}.  
The benefits of different AI assistances may depend on how much cognitive effort or cost they induce 
\citep{bucinca2021trust, vasconcelos2023explanations}, and how much humans overrely on the AI prediction. 
Recent works have compared different forms of AI assistance in different situations~\citep{bussone2015role, lai2019human, green2019principles, bansal2021does,bucinca2021trust}, as well as adapted the AI assistance to the individual~\citep{noti2022learning,ma2023who,bhatt2023learning}. 

The bulk of the work on AI assistance focusses on a single metric: accuracy.
However, in many settings, both time and accuracy are important: we also hope that the AI assistant will help us get the work done faster. 
In this paper, we specifically consider the setting where the person is under time pressure, that is, they \emph{need} to get their work done quickly and accurately (such as making decisions in an emergency room~\citep{patel2008translational,franklin2011opportunistic,rundo2020recent} or in aviation~\citep{sarter2001supporting}). 
In settings with time pressure, following the AI assistant's recommendation when the AI assistant is correct can save valuable time; in such cases, we want to encourage the person to rely on the AI assistant. 
Conversely, when the task is more difficult and the AI assistant may be incorrect, we may want to slow down the person to ensure they still make the correct decision.

Past work on AI assistance has demonstrated that different types of assistance have different accuracy-time tradeoffs.
For example, AI assistances that require more cognitive effort take longer to process, but can lead to less overreliance on the AI assistance~\citep{bucinca2021trust}. 
Conversely, AI assistances requiring less cognitive effort, such as providing an AI recommendation, can lead to increased overreliance~\citep{bussone2015role, lai2019human, jacobs2021machine}, speeding up response time but potentially lowering accuracy.  
However, these works were not in settings with explicit time pressure, and other work (that does not consider AI assistance) has shown that people may change how they perform under time pressure~\cite{kocher2006time, danielsson1999decision, klein1997recognition}.  
Understanding how people respond to AI assistance under time pressure, and what types of AI assistance is most effective, remains an open question. 

As noted above, a key element of performing tasks quickly and accurately is appropriate reliance on the AI assistant.  Thus, understanding the factors that impact the person's reliance on the AI assistant is important. 
Previous work has found that when participants are shown an AI assistance less often, they rely on it more than when they are always shown that AI assistance (`scarcity effect')~\citep{noti2022learning}. 
Because we are interested in adapting AI assistance types to get the best accuracy-time tradeoff under time pressure, we are interested in if and how the scarcity effect manifests in our setting.

In addition to the scarcity effect, different people may also have different tendencies to over- or under-rely on an AI assistant.  In our pilot studies, we asked participants how they used AI assistance after they completed their main task.  In line with prior work~\cite{sivaraman2023ignore}, we found that many people said that they either ignored the AI assistance or that they relied heavily on it. 
This indicates that perhaps people may be split into two groups based on their tendency to rely on the AI recommendation. If this is true, then different AI assistance types may benefit the two groups of people differently, impacting how we would adapt AI assistance type to the person. 

In this research, we hypothesise that participants use different AI assistance types differently under time pressure compared to not under time pressure. 
We also hypothesise that people will overrely on AI assistance more in the mixed condition (when AI assistance is withheld for some task instances and a mix of assistance types is provided on other task instances) than when always shown one type of AI assistance, due to the scarcity effect. 
Lastly, we hypothesise that there is some underlying trait that can predict whether or not a person overrelies more on the AI. 
To test these hypotheses, we conducted two experiments in a carefully-controlled setting: participants had to complete logic puzzles, with each puzzle corresponding to a sick alien that the participant needed to prescribe a medicine to. 
In both experiments, each participant either saw (i) no AI assistance, (ii) a recommendation and explanation before making a decision (`AI-before'), (iii) a recommendation and explanation after making an initial decision without assistance (`AI-after'), or (iv) a random mix of the three (`mixed'). 
AI-before is a very common AI assistance used in many settings.
AI-after has been found to improve decision accuracy compared to AI-before~\cite{green2019principles} and to reduce overreliance on AI recommendations~\citep{bucinca2021trust}. 
We introduced time pressure by having two timers on screen: one to count down the overall time remaining, and one that pressured participants to answer every puzzle within a certain amount of time. 

In our first experiment (n=159), we tested if the existence of time pressure changed a person's behaviour (their accuracy, response time and overreliance). 
We split the session into four blocks of 5 minutes each, and alternated whether the participant was shown a timer or not over the four blocks. 
We hypothesise that participants use AI assistance types differently under time pressure compared to no time pressure, and measured this through accuracy, response time and overreliance rate.
We found mixed results. 
For people who had not been under time pressure before, the introduction of time pressure led to differences in behaviour.
However, once they had been subjected to time pressure, their behaviour remained relatively stable even after removing time pressure. 
We also found we can predict a participant's tendency to overrely in the second half of the study given their overreliance behaviour on the first half of the study. 

To further test how time pressure impacts behaviour, we conducted a second experiment (n=316), where we assigned each participant to either be under time pressure or not (a between-subjects design), and used only one long 20-minute block (instead of four blocks of 5 minutes each). 
This design does not allow participants to change their behaviour after being subjected to time pressure, and mimics studies in current literature, where participants usually have no time pressure throughout the study. 
We found that the existence of time pressure impacted the accuracy-time tradeoff between different AI assistance types: under time pressure, people overrelied more on AI-before and were quicker on AI-before compared to the other conditions, which we did not see for participants that were not under time pressure. 
We also found that there was a scarcity effect under no time pressure, like in previous work \citep{noti2022learning}: AI-before assistance for participants under the mixed condition had higher overreliance and quicker response time than the participants assigned to the pure AI-before condition. 
However, under time pressure, we did not find a significant scarcity effect, indicating that the scarcity effect is not additive with time pressure: time pressure already increases overreliance, and the scarcity effect does not increase this further. 
Lastly, we found we could again predict overreliance rate in the second half of the study given overreliance behaviour in the first half, providing further evidence that there may be an underlying trait that can predict whether people overrely more. 
As a research question, we explored if we can predict a person's tendency to overrely given well-known personality traits like their Big-5 Personality Traits~\cite{gosling2003very} and Need-for-Cognition (a person's intrinsic motivation to engage in effortful mental activities)~\cite{cacioppo82:need,cacioppo96:dispositional}, and found some marginal correlations. 

As exploratory analysis, we looked into how we can adapt AI assistance type based on properties of the person (such as their tendency to overrely) and the task (such as the difficulty of the task). 
We did this by looking at participants' performance on the mixed condition, using data from our second experiment. 
We saw that the not-overrelier group achieved human-AI complementarity (with higher accuracy than both No-AI and AI-only), while the overrelier group did not. 
This further suggests that these two groups of people may benefit from different types of AI assistance under different conditions. 
We then analysed the two groups' accuracy-time tradeoff for the AI assistance types on different difficulties of questions, finding differences that we might be able to adapt to (for example, we can increase accuracy for overreliers by slowing them down with AI-after). 

In summary, we make the following contributions: 
\begin{enumerate}
    \item We expose the importance of looking at accuracy-time tradeoffs for AI-assisted decision making, instead of always focussing on finding AI assistance strategies that increase accuracy. This tradeoff is especially important to consider when using AI assistance in time-pressured scenarios, or scenarios where we hope the AI assistance will speed up human decision-making. 
    \item We find that introducing time pressure can change the accuracy-time tradeoff between different AI assistance types. Although people are quicker under time pressure for all AI assistance types, this effect is bigger for some AI assistance types than others.
    \item We find that showing different AI assistance types to the same person can impact how they use the AI assistance, like the scarcity effect in~\citet{noti2022learning}. However, we also find that this effect disappears if a person is also under time pressure. 
    \item We find evidence that there is an underlying trait that predicts a person's overreliance rate. People who overrely more are quicker on average and have different accuracy-time tradeoffs than people who overrely less, suggesting that adapting what AI assistance we show to a person's overreliance rate might be beneficial. 
    \item We provide suggestions for how we might adapt AI assistance type to participants under time pressure, keeping the accuracy-time tradeoff in mind. 
\end{enumerate}

\section{Related Work}
\label{sec:related_works}

\subsection{Decision-Making Under Time Pressure}

Research exploring the influence of time constraints on decision-making in various contexts has consistently demonstrated that individuals' decision-making abilities are compromised when subjected to time pressure~\citep{kocher2006time, danielsson1999decision, klein1997recognition}. Decision quality under time pressure often suffers because the associated stress can lead to perceptual narrowing, resulting in reduced vigilance, working memory capacity, and the utilisation of available information~\citep{klein1997recognition, sussman2022feeling, orasanu1997stress}. Studies have shown that time pressure amplifies cognitive biases~\citep{roberts2022time, diederich2020need, yik2019anchoring}, including implicit racial biases~\citep{stepanikova2012racial}. One relevant cognitive bias, known as anchoring bias, is characterised by the fact that the initial piece of information encountered, often termed as the ``anchor'', exerts an undue influence on people's decision-making~\citep{chapman1999anchoring}. 
This initial anchor can have a disproportionate influence on individuals' subsequent judgements and choices, even if the anchor is unrelated or arbitrary~\citep{englich2006playing}. People often start with an explicit or implicit anchor and through cognitive effort adjust their inference away from that anchor~\citep{epley2004perspective}. Yet when making decisions under time pressure, they may stop adjusting earlier than people who have sufficient time to make adjustments~\citep{yik2019anchoring}, often resulting in suboptimal decisions that echo the anchor. 
We believe that providing people with AI recommendations is a form of anchoring, resulting in suboptimal decisions when the AI provides incorrect recommendations, a situation which is exacerbated when people are under time pressure and have insufficient time to adjust their decisions.

Previous work has investigated the effect of individual differences on decision-making performance under various forms of pressure, such as time pressure and social pressure. Out of the Big Five Personality Traits (Openness, Conscientiousness, Extroversion, Agreeableness, Neuroticism), \citet{byrne2015chokes} find that individuals high in neuroticism and those high in agreeableness ``choke under pressure'', with their performance being severely negatively impacted by either time or social pressure compared to no-pressure situations.


\subsection{Accuracy, Reliance, and Time in AI-Assisted Decision-Making}

\textbf{Effect of different AI assistance types on overreliance.}
Initial studies expected human+AI teams to perform better than either alone \citep{kamar2012combining, amershi2019GuidelinesFH}, however, recent studies have found that this is not the case, with accuracy of the team usually worse than AI-only accuracy \citep{bussone2015role, bucinca20:proxy, green2019principles, bansal2021does, poursabzi2018manipulating, zhang2020effect}. 
This may be because humans overrely on AI predictions, making mistakes by agreeing with a wrong AI prediction (even when the human may not have made the mistake on their own), instead of achieving complementary performance \citep{bussone2015role, lai2019human, jacobs2021machine}. 
As a way to combat this, \citet{bucinca2021trust} introduced cognitive forcing functions as interaction design interventions to reduce overreliance on AI. 
They showed that the AI-after condition (or `update' condition~\citep{green2019principles}), in which participants are first asked to make a decision on their own before seeing an AI recommendation, reduced overreliance. 
But the AI-after condition may also reduce appropriate reliance, as experts may pay less attention to the recommendation after spending effort and time to make the decision unassisted \citep{fogliato2022goes}. 
We explore the AI-after condition in our work as an assistance type with the potential to reduce overreliance.

\textbf{Adaptive interventions.} To foster appropriate reliance, a few recent studies have considered adapting the AI assistance shown to users. 
\citet{noti2022learning} trained a classifier on previous data to adaptively show AI recommendations only when the AI was more likely to be correct than the human decision maker on a recidivisim prediction task. 
They found that they could increase the overall human+AI performance by showing AI recommendations only on questions that the AI was more likely to be right. 
\citet{ma2023who} also found similar results with an income prediction task. 
\citet{bhatt2023learning} considered adapting the form of AI assistance shown to different users' preferences, using contextual bandits to trade off accuracy against the cost of assistance (for example, asking an additional decision-maker for advice has higher cost than providing an ML model prediction). 
Overall, we believe these results show that adaptive interventions are a promising research direction, and we consider their potential to trade off accuracy and time.

\textbf{Predictors of response time in AI-assisted decision-making.}
To the best of our knowledge, no prior work has focussed explicitly on the tradeoff between accuracy and time in AI-assisted decision-making. 
Multiple studies, however, reported the response times of participants when shown different conditions or interventions, with mixed empirical evidence. 
Some studies found that people spent more time on instances that they perceived as more difficult inherently~\citep{arshad2015investigating, levy2021assessing}, but this additional time spent did not translate to increased accuracy. 
\citet{cao2023time} looked at the impact of time pressure on the AI-after assistance type, finding that reliance on AI recommendations can sometimes increase or decrease, depending on (i) exactly when time pressure is added during the decision-making process, and (ii) the type of task. 

For clinical annotations, \citet{levy2021assessing} found that despite additional time spent on instances with incorrect AI recommendations, human+AI accuracy was lower on those instances compared to instances with correct recommendations. 
\citet{fogliato2022goes} found that time spent on the task did not differ among AI-before and AI-after conditions. 
More related to our work, \citet{rastogi2022deciding} found that increasing the time allocated to a task alleviated anchoring bias, allowing the human decision-maker to make predictions more in line with their own knowledge as opposed to accepting the AI recommendation.

\textbf{Effect of individual differences on trust and reliance on AI.}
In the context of AI, a large number of studies have investigated the impact of individual differences on general trust and thus adoption of AI systems. A recent paper reviewed studies on how personality traits impact trust in AI across disciplines, revealing that trust in AI systems is often positively correlated with agreeableness, openness, and extroversion, but negatively correlated with neuroticism~\citep{riedl2022trust}. Note that these measures of trust are often subjective ratings rather than behavioural indicators of reliance on the AI system. 

On tasks that require cognitive effort, such as most decision-making tasks, another important personality trait is Need for Cognition (NFC), which is people's intrinsic motivation to think~\citep{cacioppo82:need}. Individuals high in NFC enjoy engaging with information, and research across domains shows that in tasks that require cognitive effort, they consistently outperform individuals low in NFC (for HCI-related examples see~\citep{carenini01:analysis,gajos17:influence,scholer2013effect}). In AI-assisted decision-making, \citet{bucinca2021trust} found that individuals high in NFC made use of AI-provided information better and benefited from cognitive forcing more than individuals low in NFC. Therefore, in addition to the Big Five personality traits, we also sought to understand whether NFC is a predictor of overreliance on AI advice.
\section{Experiment 1: Time pressure impacts behaviour}
\label{sec:experiment1}

In this first experiment, we test how time pressure impacts different AI assistance types by measuring participants' accuracy, response time and overreliance rate. 
This is a within-subject design. 
We hypothesise, 
\begin{enumerate}
    \item[\textbf{H1:}] Participants use different AI assistance types differently under time pressure compared to under no time pressure. 
\end{enumerate}

Specifically, we expect participants to overrely more under time-pressure under the AI-before condition, but not necessarily under the AI-after condition (which slows participants down, and has been shown to reduce overreliance~\citep{bucinca2021trust}), 
\begin{enumerate}
    \item[\textbf{H1.1:}] On AI-before, people overrely more under time pressure. 
\end{enumerate}

From our pilot studies, where many people either said that they ignored the AI assistance or relied on it heavily, we expect that we should be able to predict a person's overreliance rate, 
\begin{enumerate}
    \item[\textbf{H2:}] There is some underlying trait that can predict whether or not a person overrelies more on AI assistance. 
\begin{enumerate}
    \item[\textbf{H2.1:}] We can predict whether a person overrelies more or not in the second half of the study from their overreliance behaviour in the first half. 
\end{enumerate}
\end{enumerate}

\subsection{Task description}
\label{sec:expt1_task}

\begin{figure*}[t]
    \centering
    \includegraphics[width=0.9\textwidth]{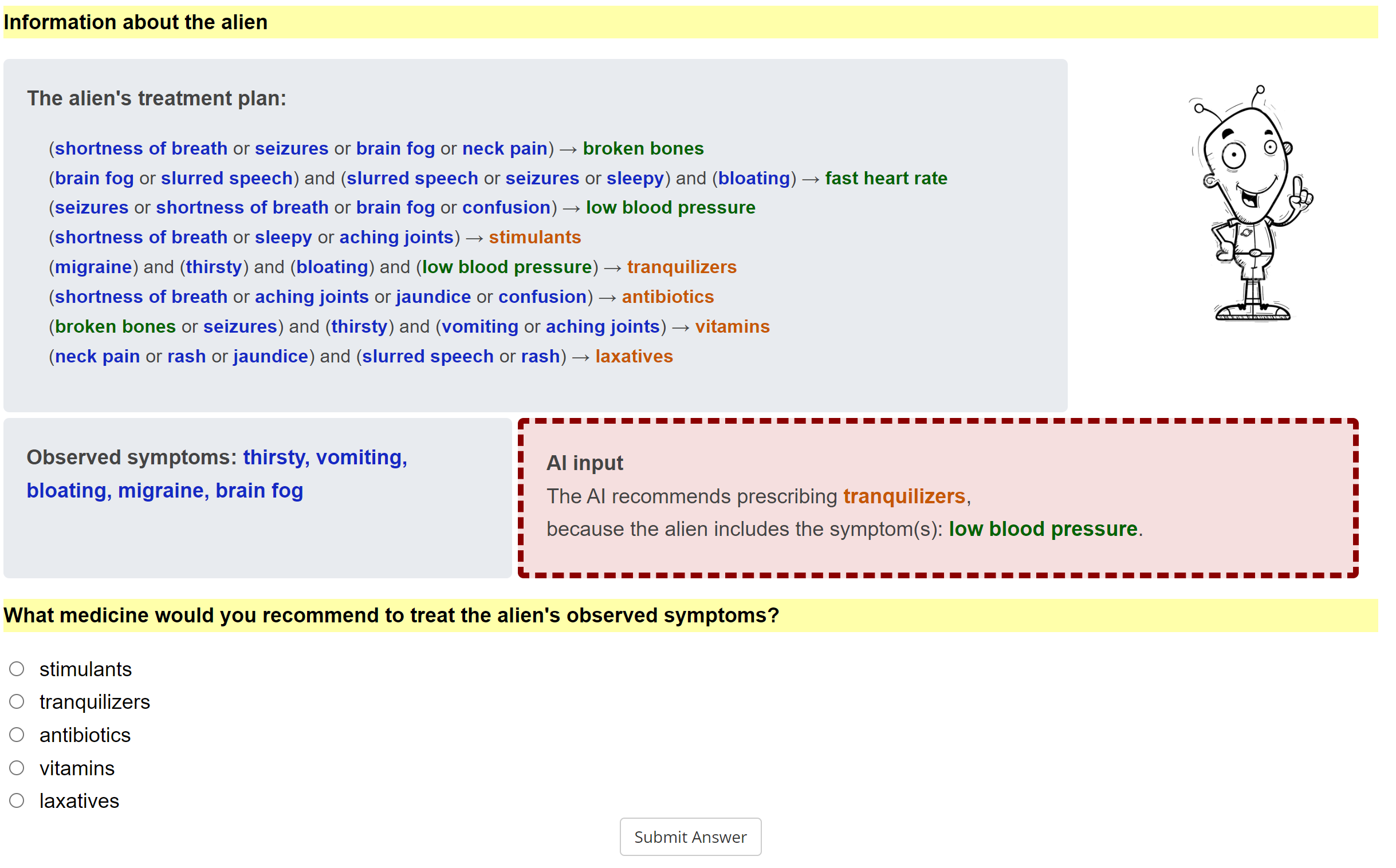}
    \caption{The alien prescription task, where participants must prescribe a single medicine. The information about the alien includes the alien's unique treatment plan (a set of rules) and the alien's observed symptoms. Participants have to use these observed symptoms and rules to prescribe a single medicine, such that only the observed symptoms and any potential intermediate (green) symptoms are used, and no other unobserved symptoms. When an AI assistance is shown, it is shown in a red box, like in this example. Here, the AI recommendation is the best possible (tranquilizers uses the most observed symptoms). 
    Vitamins is also a correct medicine, but is suboptimal as it uses fewer observed symptoms. 
    All other medicines are incorrect. 
    }
    \label{fig:alien_example}
\end{figure*}

We designed a task where users are asked to prescribe medicines to sick aliens, which we based on 
the work of~\citet{lage2019evaluation}. Our goal was to create a decision task that is accessible to laypeople but carries real-world resemblance.
Participants were shown a series of sick aliens in four `medical shifts' of 5 minutes each (with a break screen before each shift), and were asked to prescribe a single medicine to each alien. 
By asking participants to act like doctors, and by emphasising the importance of treating patients correctly, we aimed to motivate participants to obtain high accuracy, while getting through as many sick patients as possible during their medical shifts. 

\cref{fig:alien_example} shows an example of a single alien task. Based on observed symptoms and the `treatment plan' (which is a set of decision set rules unique to each alien), participants must decide a single medicine to give the alien. 
We chose to present the treatment plans as decision sets as they are relatively easy for humans to parse~\cite{lakkaraju2016interpretable}. 
When we provide an AI assistance, we show it in a red box, as shown in \cref{fig:alien_example}. 
This box provided both an AI recommendation and explanation (explanations are always an intermediate symptom that leads to the recommended medicine), and was provided before (AI-before) or after (AI-after) the participant's initial decision. 

We expanded on the setup originally introduced in~\citet{lage2019evaluation} in three ways. 
First, we always introduced intermediate symptoms to the task, which required participants to perform additional computation steps, and worked well as the explanation of an AI's recommendation.
Second, we allowed two possible correct medicines per alien. 
We defined the better medicine to be one that addressed more of the observed symptoms. 
Having a suboptimal medicine helped us to better analyse the role of overreliance on AI recommendations: suboptimal medicines could easily be verified to be correct (even though they were not optimal), and so participants could overrely on them more easily than overrelying on a wrong recommendation. 

Third, we introduced two different levels of difficulty of questions: easy and hard. 
We designed these such that easy questions required less cognitive effort for a human to find the best medicine, while hard questions required more computation. 
We ensured that both easy and hard questions superficially appeared very similar to a human, by having a similar length of lines, number of lines, and other visual aspects. 
\cref{fig:alien_example} is an example of an easy question, while \cref{fig:alien_example_hard_timer} (in \cref{app:1}) is a hard question. 

While the task is in a fictional setting, it allows us to precisely manipulate the difficulty of the task, the optimality of the AI assistance, and the form of time pressure. This allows us to understand how these factors impact human decision-making in settings where they must complete a large number of tasks in a short time.

\subsection{Conditions}
\label{sec:expt1_conditions}

Our study was a mixed between- and within-subjects design. The between-subject factor was the condition (\textit{No-AI}, \textit{AI-before}, \textit{AI-after}, \textit{Mixed}), and the within-subject factor was time pressure (\textit{time pressure} and \textit{no time pressure}).

We considered four conditions, and randomly assigned participants to one of them.
\begin{enumerate}
    \item \textit{No-AI}: Do not provide any AI assistance. 
    \item \textit{AI-before}: An AI recommendation and explanation is provided to the participant along with the question, before the participant makes any decision.
    \item \textit{AI-after (or, `update')}: The participant makes an initial decision without any AI assistance. They are then provided with an AI recommendation and explanation, and allowed to change their initial answer. 
    \item \textit{Mixed}: For every question, the participant randomly gets one of the three above assistance types (No-AI, AI-before, AI-after). 
\end{enumerate}

We assigned half the participants to the first (No-AI) condition compared to each of the other three AI-assistance conditions.
This is because we primarily used the No-AI condition to see if participants achieve human-AI complementarity (if accuracy with an AI assistance is greater than both No-AI accuracy and AI-only accuracy). This required fewer participants than our detailed comparisons between the other three AI-assistance conditions. 

We alternated whether or not there was time pressure (whether or not there were timers shown on the screen) across each participant's four shifts (the blocks of 5 minutes). 
We randomly assigned participants to either have time pressure or no time pressure in the first shift. 

For the shifts without a timer, participants saw the screen in~\cref{fig:alien_example}. 
For the shifts with time pressure, participants saw two timers on the screen: a global timer and a local timer (see \cref{fig:alien_example_hard_timer} in \cref{app:1} for an example). 
The global timer indicated the time remaining for participants to answer questions in their shift (the timer started at 5 minutes and counted down). 
The local timer was reset to 60 seconds at the beginning of every question, and signified the `recommended time' remaining to treat the specific alien patient. When it reached 10 seconds, the timer's text colour turned orange, and when it reached 0 seconds, the colour turned red (and the timer went into negative numbers). 
After participants provided their initial response in the AI-after assistance type, we set the timer to 20 seconds (the recommended time they had to update their initial answer given the AI input). 
Although there was no requirement for participants to answer each question within the local timer's 60 seconds, the presence of the timer increased time pressure on the participant. 
The local timer allowed us to impose a type of time pressure that applies evenly to all parts of their shift, rather than participants feeling relaxed during the first part of their shift and increasingly pressured toward the end of their shift.

\subsection{Procedure}
\label{sec:expt1_procedure}

Our study was conducted online on Prolific, a crowdsourcing platform. Before starting the main part of the study, participants had to accept a consent form, and then answer three pages of survey questions.
The first page asked demographic information, the second page asked 4 Need For Cognition trait questions (the same subset of the original questionnaire~\citep{cacioppo84:efficient} as used by~\citet{gajos17:influence}) and 2 questions about performance under time pressure~\citep{buser2022time}, and the third page asked questions about their Big-5 Personality Traits (the first 10 questions were the BFI-10 questions~\citep{rammstedt2007measuring}, and last two were the two questions about neuroticism from TIPI~\citep{gosling2003very,kankaras2017personality}). 
Participants then had to read instructions, and then successfully complete three practice questions (for which they had two attempts, similar to~\citet{lage2019evaluation}: if they failed the first set of practice questions, they were given a second set). 
Participants were provided feedback on how to improve their answer during the practice questions, but not during the main study. 

After completing the practice questions, the participants saw a screen telling them that their practice was over, and they then started their first shift of 5 minutes, during which they had to answer as many questions as possible. 
They had four shifts of 5 minutes each, with a break screen in between every shift. 
Approximately half of the questions participants saw were easy questions, and the other half hard questions. 
The order of difficulty was randomised (every question had a 50\% chance of being easy or hard). 

After the study, participants were shown a final screen where they were asked (i) how difficult they found the task (on a 5-point scale from `Very easy' to `Very difficult'), (ii) how helpful they found the AI assistance (on a 5-point scale from `Very unhelpful' to `Very helpful'), and three open-ended questions asking (iii) what their strategy for approaching the task was, (iv) if their strategy changed when there was an AI input, and (v) for any other feedback. 

We ran the study with 207 participants on Prolific. 
Only English speakers in the US were allowed to participate. 
44 people failed the practice questions, and we removed a further 4 people for either answering questions too quickly (they spent less than 3 seconds on at least 3 questions), or taking much longer on one question than the rest (indicating they were distracted for one question). 
Our final results are based on the remaining 159 participants. 

Participants' mean age was 39 years (standard deviation of 13 years). 
79 participants self-identified as male, 76 as female, 2 as non-binary, and 2 preferred not to say their gender. 47 of the participants' highest level of education was high school, 74 had a bachelor's degree, 31 had Master's (or beyond), and 7 answered `other' for highest level of education. 

Participants were paid \$7 (US) for participating (median time was 38 minutes, corresponding to $\$11.05$/hr), and if they failed the practice questions, they were paid \$2 (their study ended immediately after failing the practice questions). 
We also incentivised participant performance by providing a bonus \$3 reward to the top-performing participant in each condition. 

\textit{Approvals.} Both experiments in our paper were approved by the Internal Review Board at Harvard University, protocol number IRB15-2076. 

\subsection{Design and analysis}
\label{sec:expt1_analysis}

We report three metrics.
\begin{enumerate}
    \item \textit{Accuracy}: if participants chose the \textit{best} medicine for the alien, we gave them a score of 1, a \textit{suboptimal} (but correct) medicine has a score of 0.5, and a \textit{wrong} medicine has a score of 0. We calculate the average accuracy over questions for each participant, and report mean and standard error across participants. 
    \item \textit{Response time}: we measure how long each participant takes to answer questions. We report the mean and standard error across participants. 
    \item \textit{Overreliance}: we define overreliance to be the proportion of times a participant gave the same answer as the AI when the AI was wrong or suboptimal \citep{bucinca2021trust, vasconcelos2023explanations}. 
\end{enumerate}

We fixed the AI to have an average accuracy of 0.70. For every question, there was a 60\% chance the AI recommended the best medicine, 20\% chance of suboptimal medicine, and 20\% chance of a wrong medicine. 
We additionally ensured that the AI recommended the best medicine for the first two recommendations, as previous work found that the first few interactions with an AI can bias a participant to distrust the AI if the AI gives wrong recommendations~\cite{nourani2021anchoring}. 
We adjusted the AI's chances of recommending best/suboptimal/wrong medicines such that overall AI accuracy remained at 0.70, despite recommending the best medicine for the first two questions. 

We used analysis of variance to compare across our AI assistance types on our three metrics, and then Tukey's HSD for post-hoc pairwise comparisons. 

When comparing within-subject (such as how time pressure impacts a participant's behaviour), we performed pairwise tests (comparing how time pressure changed performance for each of the AI-assisted conditions) using the Holm-Bonferroni correction method~\citep{holm79:simple} to correct for repeating this analysis on each of the three conditions. 

We split participants into two equal groups (overreliers and not-overreliers) based on their overreliance rate. 
When predicting if a participant is an overrelier or not later in the study given whether they were an overrelier earlier in the study, we used a logistic regression model and ran a $\chi^2$ test.

\subsection{Results}
\label{sec:expt1_results}

\subsubsection{Time pressure impacts how participants use AI assistance types differently (hypothesis H1)} 

Overall, we find some limited evidence that participants use different AI assistance types differently under time pressure compared to no time pressure (hypothesis H1). 
Performance metrics are summarised in \cref{table:expt1_results}. 
Under no time pressure, all AI-assisted conditions have similar accuracies ($F(2, 136)=1.11, p=0.33$).
However, under time pressure, we observed a significant main effect of AI condition on accuracy ($F(2, 136)=5.99, p=0.003$). 
Specifically, mixed had a lower accuracy than AI-before and AI-after ($p=0.044$ and $p=0.003$ respectively).
There is also a significant main effect of AI condition on response time under time pressure ($F(2, 136)=10.3, p<0.0001$), where mixed is also now faster than AI-after ($p=0.006$). 
Under both time pressure and no time pressure, AI-before is faster than AI-after ($p=0.018$ under no time pressure, and $p<0.0001$ under time pressure). 

Both under no time pressure and under time pressure, we observed a significant main effect of AI condition on overreliance (under no time pressure $F(2, 134)=4.89, p=0.0089$, and under time pressure $F(2, 136)=5.41, p=0.0055$). Specifically, both with and without time pressure participants overrelied on mixed more than AI-after with (no time pressure: $p=0.007$, with time pressure: $p=0.004$). 

We do not find that people overrely more on AI-before under time pressure compared to no time pressure (hypothesis H1.1): $p=0.13$. 

\begin{table*}[t]
    \centering
    \begin{tabular}{lcccccc}
        \toprule 
        \textbf{} & \multicolumn{3}{c}{--- \textbf{No time pressure} ---}  & \multicolumn{3}{c}{--- \textbf{Time pressure} ---} \\     
        \textbf{Condition} & \textbf{Acc} & \textbf{Time (s)} & \textbf{Overreliance} & \textbf{Acc} & \textbf{Time (s)} & \textbf{Overreliance} \\
        \hline
        
        \textbf{No-AI}      & 0.58(0.05) & 63(11) & ---       & 0.59(0.04) & 41(5) & ---  \\
        \textbf{AI-before}  & 0.74(0.02) & 48(4) & 0.51(0.05) & 0.72(0.02) & 39(3) & 0.56(0.05)  \\
        \textbf{AI-after}   & 0.75(0.02) & 65(4) & 0.37(0.04) & 0.75(0.02) & 56(3) & 0.42(0.05)  \\
        \textbf{Mixed}      & 0.70(0.02) & 61(4) & 0.59(0.06) & 0.65(0.02) & 44(2) & 0.65(0.05)  \\

        \bottomrule
    \end{tabular}
        \caption{Mean (standard error in parentheses) for our three metrics, averaged over the shifts that participants are not under time pressure (left) and under time pressure (right). 
        We find some limited evidence that participants use different AI assistance types differently in the two settings. 
        Under no time pressure, all AI-assisted conditions have similar accuracy, while AI-before is quicker and has higher overreliance. 
        Under time pressure, Mixed has lower accuracy than the others, and both AI-before and Mixed are quicker and have higher overreliance than AI-after. 
        See text for details on statistical analysis.}
    \label{table:expt1_results}
\end{table*}

When we look at these results in more detail, we find that participants who did not have time pressure in the first block behaved differently on the first block compared to their second no-time-pressure block. 
In particular, by the time they approached their second no-time-pressure block, they had already completed a block under time pressure, and they performed similarly under no-time-pressure as they did under time pressure (such as significantly faster response times). 
This also likely affected participants who saw a time-pressure block first: their performance under no-time-pressure was more similar to the time-pressure blocks. 
We therefore designed a second experiment (see \cref{sec:experiment2}), where participants were either assigned to no-time-pressure or to time-pressure only. 
This is more similar to studies in current literature, where participants usually have no time pressure throughout the study, and so cannot adapt their behaviour after experiencing the same task under time pressure.

\subsubsection{We can predict a participant's overreliance rate (hypothesis H2)}
\label{sec:expt1_results_overreliance}

We find we can predict whether or not a participant was an overrelier in the second half of the study (the last two shifts) given whether or not they were an overrelier in the first half of the study (the first two shifts) ($\chi^2(1, N=93)=16.28, p<0.0001$), confirming hypothesis H2.1 and providing evidence towards hypothesis H2. 

Given our observation earlier that participants' behaviour was different before they saw any time pressure, we repeat the above analysis, this time trying to predict whether a participant was an overrelier on the last three shifts given whether or not they were an overrelier on the first shift. When the first shift was a time-pressure shift, we can predict successfully ($\chi^2(1,N=59)=7.47, p=0.0063$). When the first shift was a no-time-pressure shift, we cannot predict their overreliance group ($\chi^2(1,N=41)=0.22, p=0.64$). 
This further indicates that a participant's behaviour changes after experiencing time pressure.

\section{Experiment 2: Assigning different participants to time pressure or no time pressure}
\label{sec:experiment2}

In the previous experiment, we found some evidence that people use AI assistance differently under time pressure compared to no time pressure, but we found that a participant's behaviour sticks after completing the task under time pressure, affecting how they behave under no time pressure later. 
We therefore design a second experiment, where each participant was only assigned to a no-time-pressure condition or time-pressure condition. 

We make the same hypothesis regarding behaviour under time pressure, 
\begin{enumerate}
    \item[\textbf{H1:}] Participants use different AI assistance types differently under time pressure compared to under no time pressure. 
    \begin{enumerate}
        \item[\textbf{H1.1:}] On AI-before, people overrely more under time pressure. 
    \end{enumerate}
\end{enumerate}

As in experiment 1, we expect that we should be able to predict a participant's overreliance rate, 
\begin{enumerate}
    \item[\textbf{H2:}] There is some underlying trait that can predict whether or not a person overrelies more on AI assistance. 
\begin{enumerate}
    \item[\textbf{H2.1:}] We can predict whether a person overrelies more or not in the second half of the study from their overreliance behaviour in the first half. 
\end{enumerate}
\end{enumerate}

We also look in more detail at the mixed condition. We hypothesise to see the scarcity effect \citep{noti2022learning} in the mixed condition, and we measure this through overreliance rate on the AI-before assistance type. 
Overreliance on AI-after may not increase, as the purpose of AI-after is to slow people down and reduce overreliance, and so we do not hypothesise about it.  
\begin{enumerate}
    \item[\textbf{H3:}] In the mixed condition, overreliance rate on AI-before is higher than for the pure AI-before condition. 
    \begin{enumerate}
        \item[\textbf{H3.1:}] Overreliance rate is higher under no time pressure. 
        \item[\textbf{H3.2:}] Overreliance rate is higher under time pressure. 
    \end{enumerate}
\end{enumerate}

We also ask, as a research question, if we can predict a participant's overreliance using personality traits that we estimate by asking questions at the beginning of the study,
\begin{enumerate}
    \item[\textbf{RQ1:}] Can we use personality traits to predict whether or not a person overrelies more on AI assistance?
\end{enumerate}

\subsection{Task description and conditions}
\label{sec:expt2_task_conditions}

The task design was identical to experiment 1 (see \cref{sec:expt1_task}). However, this study was a between-subject design only, with AI-condition (4 options) and time pressure (2 options) both as the between-subject factors (8 conditions in total). AI-condition again consisted of four levels (No-AI, AI-before, AI-after, mixed), same as in Experiment 1. 
As before, we assigned half the number of participants to the first (No-AI) condition compared to each of the other conditions. 

The key difference compared to experiment 1 is that time pressure was a between-subject rather than within-subject factor in this experiment. Participants now answered questions in a single block of 20 minutes, instead of four blocks of 5 minutes each. 
Correspondingly, each participant was assigned to either be under no time pressure (no timers on the screen), or time pressure (they see timers shown on screen). 

\subsection{Procedure}
\label{sec:expt2_procedure}

As in experiment 1, our study was conducted online on Prolific. The procedure was the same as in experiment 1 (except now there is just one shift of 20 minutes). 

We ran the study with 403 participants on Prolific. 
Only English speakers in the US were allowed to participate. 
75 people failed the practice questions, and we removed a further 12 people for either answering questions too quickly (they spent less than 3 seconds on at least 3 questions), or taking much longer on one question than the rest (indicating they got distracted for one question). 
Our final results are based on the remaining 316 participants. 

Participants' mean age was 39 years (standard deviation of 13 years). 161 participants self-identified as male, 147 as female, 6 as non-binary, and 2 preferred not to say their gender. 111 of the participants' highest level of education was high school, 145 had a bachelor's degree, 40 had Master's (or beyond), and 20 answered `other' for highest level of education. 

Participants were paid \$7 for participating (median time was 37 minutes, corresponding to $\$11.35$/hr), and if they failed the practice questions, they were paid \$2 (their study ended immediately after failing the practice questions). 
We also incentivised participant performance by providing a bonus \$3 reward to the top-performing participant in each condition.

\subsection{Design and analysis}
\label{sec:expt2_analysis}

We report the same three metrics as in \cref{sec:expt1_analysis}: accuracy, response time and overreliance. 

As in \cref{sec:expt1_analysis}, we used analysis of variance to compare across our AI assistance types on our three metrics, and then Tukey's HSD for post-hoc pairwise comparisons. 

To compare between time pressure and no time pressure conditions on a specific condition, we performed a pairwise test for each condition, correcting using the Holm-Bonferroni method (to correct for the multiple conditions). 
When predicting if a participant is an overrelier or not later in the study given whether they were an overrelier earlier in the study, we used a logistic regression model and ran a $\chi^2$ test.

\subsection{Results}
\label{sec:expt2_results}

\subsubsection{Time pressure impacts how participants use AI assistance types differently (hypothesis H1)}

\begin{figure*}[t]
     \centering
     \includegraphics[width=0.6\textwidth]{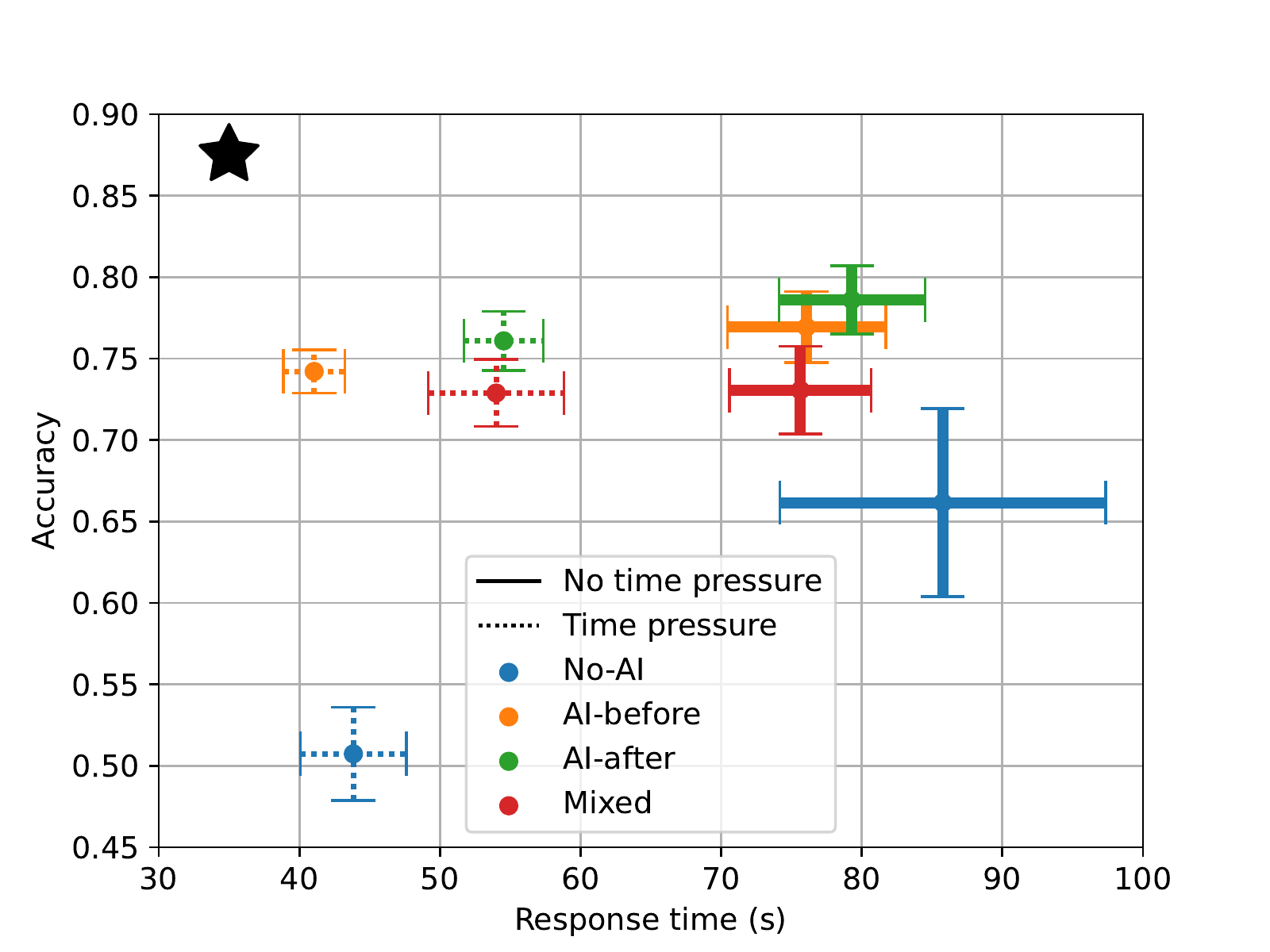}
     \caption{
     Pareto plots: top left is desired (higher accuracy, lower response time per question). We plot the performance (mean and standard error) of our four conditions (No-AI, AI-before, AI-after and mixed) under no time pressure (bold lines) and under time pressure (dotted lines). 
     We see that, under time pressure, all conditions become quicker, but AI-before becomes much quicker than the others, while keeping similar accuracy as the other AI assistance conditions. 
     No-AI reduces accuracy under time pressure. 
     See \cref{table:expt2_results} for values. 
     }
     \label{fig:pareto_effect_of_pressure}
\end{figure*}

\begin{table*}[t]
    \centering
    \begin{tabular}{lcccccc}
        \toprule 
        \textbf{} & \multicolumn{3}{c}{--- \textbf{No time pressure} ---}  & \multicolumn{3}{c}{--- \textbf{Time pressure} ---} \\        
        \textbf{Condition} & \textbf{Acc} & \textbf{Time (s)} & \textbf{Overreliance} & \textbf{Acc} & \textbf{Time (s)} & \textbf{Overreliance} \\
        \hline
        
        \textbf{No-AI}      & 0.66(0.06) & 86(12) & ---       & 0.51(0.03) & 44(4) & ---  \\
        \textbf{AI-before}  & 0.77(0.02) & 76(6) & 0.41(0.04) & 0.74(0.01) & 41(2) & 0.59(0.04)  \\
        \textbf{AI-after}   & 0.79(0.02) & 79(5) & 0.36(0.04) & 0.76(0.02) & 55(3) & 0.44(0.05)  \\
        \textbf{Mixed}      & 0.73(0.03) & 76(5) & 0.50(0.06) & 0.73(0.02) & 54(5) & 0.44(0.05)  \\

        \bottomrule
    \end{tabular}
        \caption{Mean (standard error in parentheses) for our three metrics on Experiment 2. 
        Under no time pressure, the AI-assisted conditions have similar accuracy, response time and overreliance. 
        Under time pressure, AI-before is quicker and has higher overreliance than the others, while all still have similar accuracy. 
        Also see \cref{fig:pareto_effect_of_pressure} to visualise the accuracy-time tradeoffs. 
        See text for details on statistical analysis. }
    \label{table:expt2_results}
\end{table*}

The effect of time pressure on the conditions is summarised in \cref{fig:pareto_effect_of_pressure} and \cref{table:expt2_results}. 
We see that all conditions speed up under time pressure. 
Under no time pressure, AI-before, AI-after and mixed have similar accuracy ($F(2, 135)=1.46, p=0.24$), response time ($F(2, 135)=0.15, p=0.86$) and overreliance ($F(2, 135)=2.37, p=0.10$). 
However, under time pressure, we observed a significant main effect of AI condition on response time ($F(2, 135)=5.09, p=0.0073$) and overreliance ($F(2, 135)=4.09, p=0.019$). 
AI-before is now quicker than AI-after and mixed ($p=0.02$ for both), and has higher overreliance rate than AI-after and mixed ($p=0.04$ for both), while all AI-assisted conditions still have similar accuracy ($F(2, 135)=0.80, p=0.45$). 
This indicates that, under time pressure, AI-before is better for the accuracy-time tradeoff, while this was not the case under no time pressure (all assistance types were equally good). 
This confirms our hypothesis H1. 

We also see that overreliance rate for AI-before increases significantly under time pressure ($p=0.008$), confirming hypothesis H1.1. 

\subsubsection{We can predict a participant's overreliance rate (hypothesis H2)}

Like in experiment 1 (see \cref{sec:expt1_results_overreliance}), we find that we can predict whether or not a participant is an overrelier in the second half of their block (last 10 minutes) given whether or not they are an overrelier in the first half (first 10 minutes), both under no time pressure ($\chi^2(1,N=130)=19.6, p<0.0001$) and under time pressure ($\chi^2(1,N=136)=16.1, p<0.0001$). 
This confirms hypothesis H2.1, and indicates that a participant's overreliance behaviour remains stable over the course of study. 

We also see if overreliance and response time are related, expecting that overreliers are quicker to answer questions, as further evidence that overreliance may be stable for a participant.
We find that we can predict whether or not a participant is an overrelier given their average response time (we test if there is sufficient evidence for Pearson's correlation coefficient to not be zero, finding overreliance rate response time are negatively correlated, $r(136)=-0.04, p=0.0008$ under time pressure, and $r(136)=-0.02, p<0.0001$ under no time pressure.

\subsubsection{There is a scarcity effect under no time pressure, but not under time pressure (hypothesis H3)}

\begin{figure*}[t]
     \begin{subfigure}{0.49\textwidth}
         \centering
         \includegraphics[width=\textwidth]{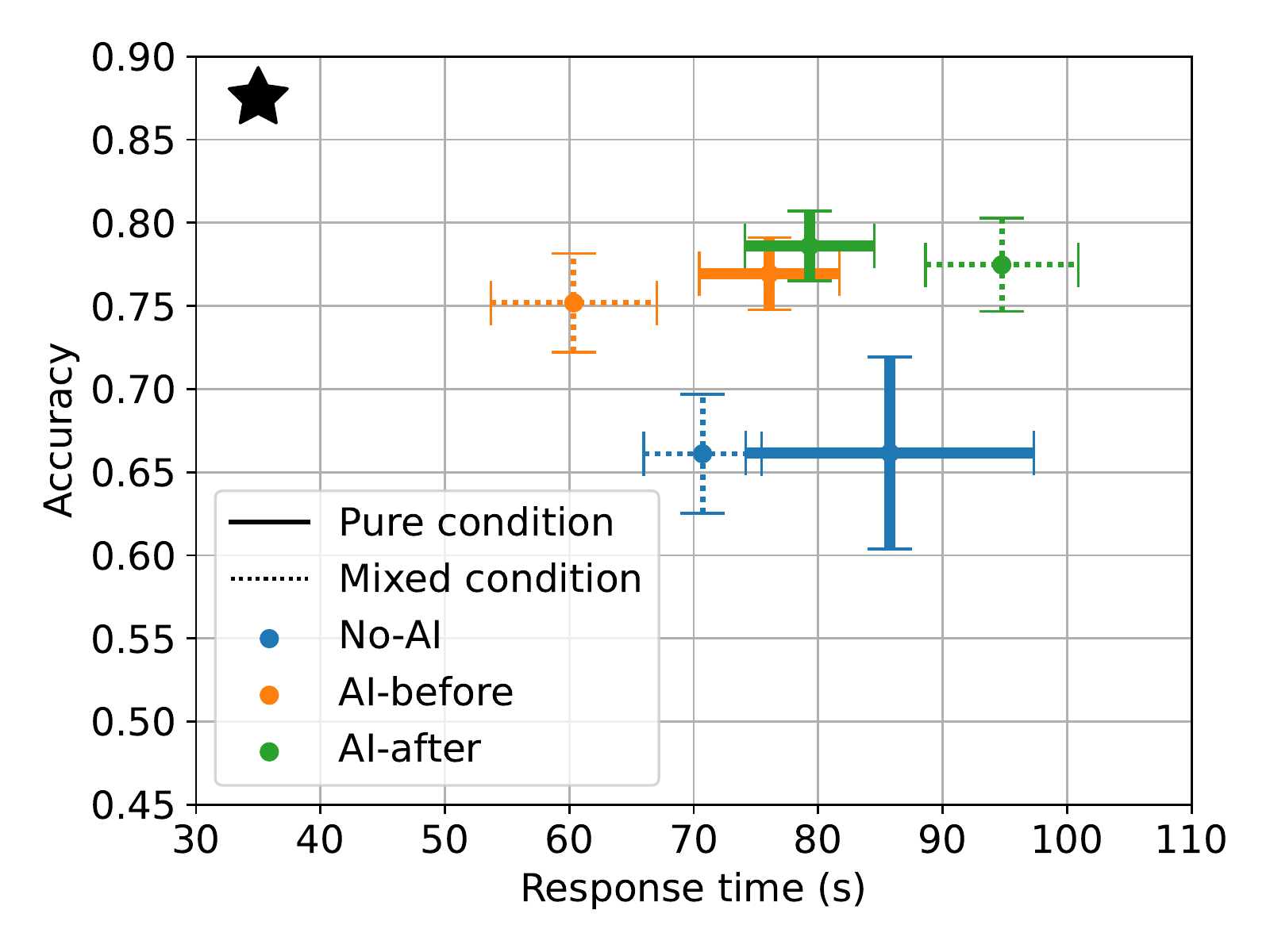}
         \caption{Under no time pressure.}
         \label{fig:pareto_no_time_pressure_mixed}
     \end{subfigure}
     \begin{subfigure}{0.49\textwidth}
         \centering
         \includegraphics[width=\textwidth]{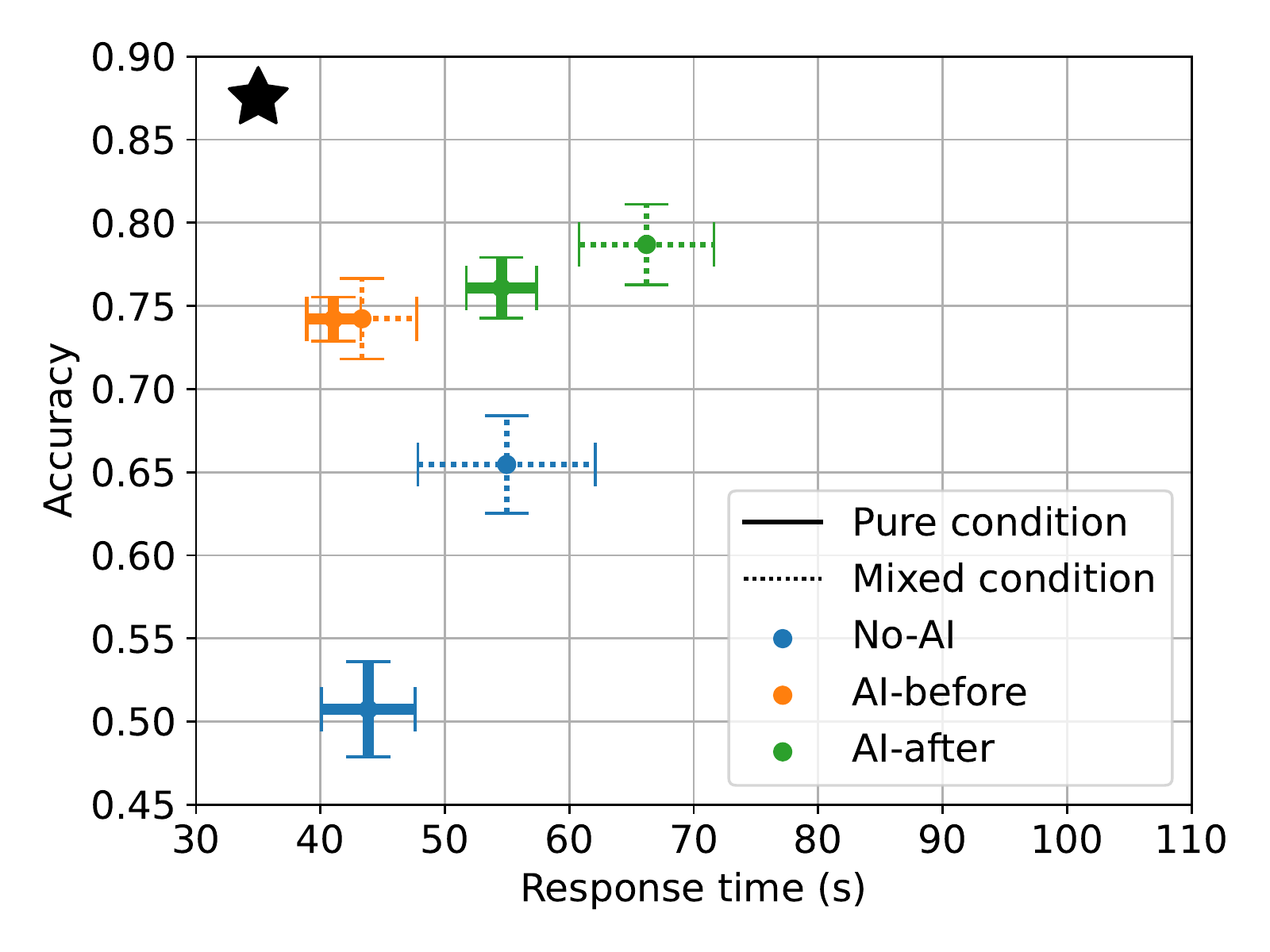}
         \caption{Under time pressure.}
         \label{fig:pareto_time_pressure_mixed}
     \end{subfigure}
     \caption{
     Pareto plots: top left is desired (higher accuracy, lower response time per question). 
     We compare the performance (mean and standard error) of the AI assistances in the mixed condition (dotted lines) to the pure AI assistance-only conditions (bold lines). 
     We see that, under no time pressure (left), mixed AI-before is marginally faster than pure AI-before (with significantly higher overreliance, see main text), and mixed AI-after is marginally slower than pure AI-after. 
     Under time pressure (right), mixed AI-before now has the same response time as pure AI-before, and mixed No-AI has higher accuracy than pure No-AI.     
     }
     \label{fig:pareto_effect_of_pressure_mixed}
\end{figure*}

The performance of AI assistances under the mixed condition are summarised in \cref{fig:pareto_effect_of_pressure_mixed}. 
We find that, under no time pressure, people overrely on mixed AI-before significantly more than pure AI-before ($p=0.001$).
This confirms hypothesis H3.1, and is the scarcity effect as observed in \citet{noti2022learning}. 

However, under time pressure, people do not overrely on mixed AI-before more than pure AI-before, meaning we do not find evidence to support hypothesis H3.2. 
This indicates that the scarcity effect is not additive with time pressure: time pressure already increases overreliance, and the scarcity effect does not increase this further. 

Looking into the accuracy-time tradeoff, we see that mixed AI-before is always significantly faster than mixed AI-after. Under no time pressure, mixed AI-before's response time of $60(7)$ seconds (standard error reported in parentheses) is quicker than AI-after's $95(6)$ seconds ($p<0.0001$). 
Under time pressure, mixed AI-before's response time of $43(4)$ seconds is quicker than AI-after's $66(5)$ seconds ($p=0.017$).

\subsubsection{We can only marginally predict overreliance from personality traits (RQ1)}

Given that we have evidence for overreliance being stable for a participant, we see if we can predict their overreliance behaviour from well-known personality traits. 
We try to predict whether or not a participant is an overrelier or not based on the Big-5 Personality Traits, Need-for-Cognition (NFC) trait, and self-reported time-pressure performance (see \cref{sec:expt1_task} for more details on these traits), and test if Pearson's correlation coefficient is not zero. 
These personality traits are estimated based on questions we asked at the beginning of the study. 
We do not find any significant results, but do find some traits that can predict overreliance with marginal significance. 
Under no time pressure, neuroticism and NFC are marginally negatively correlated with predicting overreliance group ($r(136)=-0.28, p=0.07$ for neuroticism, $r(136)=-0.31, p=0.10$ for NFC). 
Under time pressure, neuroticism is still marginally negatively correlated ($r(136)=-0.23, p=0.16$), and time-pressure-score is marginally positively correlated ($r(136)=0.35, p=0.051$). 
Future work is required to test if these traits (or a combination of them) are predictive of overreliance behaviour. 

\section{Exploratory Analysis: Adapting AI assistance to the person and task}
\label{sec:exploratory-adapting}

In this section, we use the data from Experiment 2 (\cref{sec:experiment2}) to explore how we might adapt AI assistance depending on both the person (whether they are an overrelier or not) and the task (the difficulty of the question, which is either easy or hard). 

We first split people into two equal groups (overreliers and not-overreliers) in \cref{sec:exploratory-complementarity}, and find that not-overreliers achieve human-AI complementarity (higher accuracy than both No-AI accuracy and AI-only accuracy), while overreliers do not. 
This highlights the usefulness of splitting people into these two groups: the two use AI assistance differently (one to achieve complementarity, while the other overrelies on it). Therefore adapting AI assistance to these groups should be useful. 
We then see if, after splitting people by their overreliance behaviour in the first half of the study, their accuracy in the second half of the study achieves human-AI complementarity. 
We again find that not-overreliers achieve complementarity, while overreliers do not, indicating that overreliance rate (and its corresponding behaviour) is stable during the study. 

We then explore adapting to both the person (whether they are an overrelier or not) and the question (whether the question is easy or hard) in \cref{sec:exploratory-adapting-time-pressure}, focussing on the setting where people are under time pressure. 
We find that we can improve accuracy by slowing down overreliers with AI-after on hard questions. 
For not-overreliers however, AI-before is usually faster than AI-after, with similar accuracy.  

\subsection{Some people achieve human-AI complementarity, but others do not}
\label{sec:exploratory-complementarity}

In this section, we compare participants' accuracy with AI assistance against No-AI accuracy and AI-only accuracy, to see if participants achieve human-AI complementarity (greater accuracy than both No-AI and AI-only). 
We find that, when we split the participants into two groups, overreliers and not-overreliers, the not-overreliers achieve human-AI complementarity (both under time pressure and not), while the overreliers do not. 

Results are summarised in \cref{table:complementarity}. When comparing against No-AI, we first use analysis of variance, and if there is significance, we perform pairwise comparisons using Tukey's HSD. 
When comparing against AI-only accuracy, we conduct a within-subject t-test, as we know the AI-only accuracy for each participant, and correct using the the Holm-Bonferroni method for the three conditions. 

\begin{table*}[th]
    \centering
    \begin{tabular}{llcccc}
        \toprule 
        \textbf{} & \textbf{} & \multicolumn{2}{c}{\textbf{No time pressure}}  & \multicolumn{2}{c}{\textbf{Time pressure}} \\        
        \textbf{} & \textbf{Condition} & \textbf{> No-AI acc?} & \textbf{> AI acc?} & \textbf{> No-AI acc?} & \textbf{> AI acc?} \\
        \hline
        Overreliers     & & {\footnotesize $F_{3, 81}=2.25, p=.089$}  & & {\footnotesize $F_{3, 91}=22.4, p<.0001$} & \\
                        & AI-before & \textit{n.s.} & \textit{n.s.} & Yes ($p<0.0001$) & \textit{n.s.}  \\
                        & AI-after  & \textit{n.s.} & \textit{n.s.} & Yes ($p<0.0001$) & \textit{n.s.} \\
                        & Mixed     & \textit{n.s.} & Yes ($p=0.0061$) & Yes ($p<0.0001$) & \textit{n.s.} \\
        \hline
        Not-overreliers     & & {\footnotesize $F_{3, 83}=5.68, p=.0014$}  & & {\footnotesize $F_{3, 89}=27.5, p<.0001$} & \\
                        & AI-before & Yes ($p=0.041$) & Yes ($p<0.0001$) & Yes ($p<0.0001$) & Yes ($p=0.0023$)  \\
                        & AI-after  & Yes ($p=0.001$) & Yes ($p<0.0001$) & Yes ($p<0.0001$) & Yes ($p=0.0035$)  \\
                        & Mixed     & Yes ($p=0.012$) & Yes ($p<0.0001$) & Yes ($p<0.0001$) & Yes ($p=0.0035$)  \\
        \bottomrule
    \end{tabular}
        \caption{Comparing accuracy of the AI assistance conditions with No-AI accuracy and AI-only accuracy after splitting participants by overreliance. 
        We calculate all accuracies and whether a participant is an overrelier or not over the course of the entire 20-minute study. 
        We see that, both under no time pressure and under time pressure, overreliers do not significantly achieve human-AI complementarity (accuracy is not significantly higher than both No-AI accuracy and AI-only accuracy), while not-overreliers do achieve complementarity.}
    \label{table:complementarity}
\end{table*}

Our findings indicate that we should consider adapting what AI assistance we show depending on whether the participant is an overrelier or not, as these two groups use AI assistances in different ways. 
For example, overreliers may benefit from AI assistance types that slow them down and force them to engage more with the AI. 
We investigate this further in our setting in \cref{sec:exploratory-adapting-time-pressure}. 

We earlier saw that we can predict whether a participant is an overrelier or not in the second half of the study from whether or not they were an overrelier in the first half (\cref{sec:expt1_results,sec:expt2_results}), indicating that their overreliance behaviour is stable during the study. 
We now see if, after classifying people as overreliers or not-overreliers based on the first half of the study, not-overreliers still achieve human-AI complementarity in the second half of the study, while overreliers do not. 
Results are summarised in \cref{table:complementarity_split_half} (in \cref{app:1}): not-overreliers achieve complementarity (except that, under no time pressure, the AI-before condition now has only marginally higher accuracy than No-AI ($p=0.06$)), while overreliers do not achieve complementarity. 
This provides further evidence that overreliance rate (and its corresponding behaviour) is stable during the study.

\subsection{Adapting AI assistance to both the person and task under time pressure}
\label{sec:exploratory-adapting-time-pressure}

\begin{figure*}[t]
     \begin{subfigure}{0.49\textwidth}
         \centering
         \includegraphics[width=\textwidth]{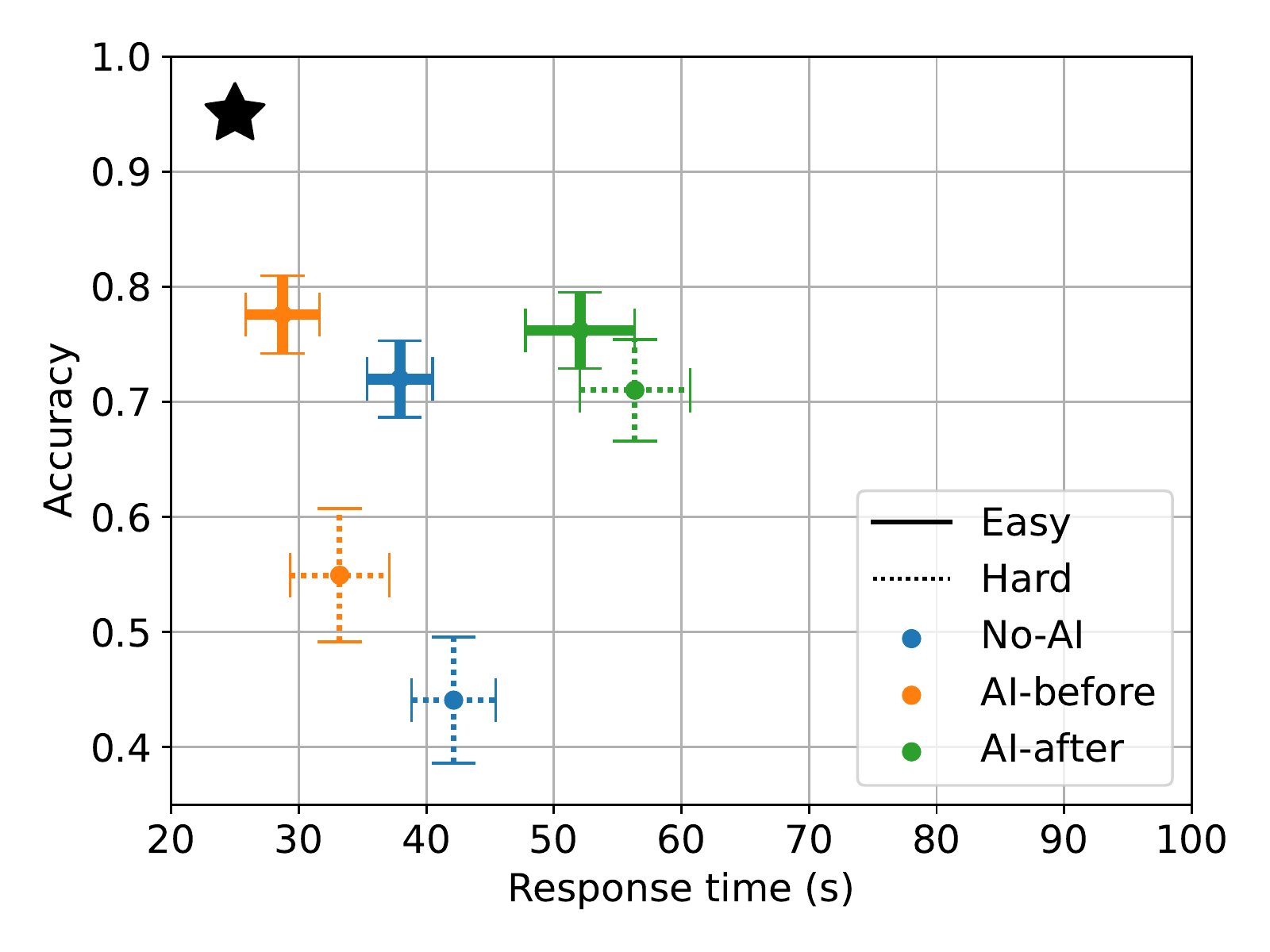}
         \caption{Overreliers.}
         \label{fig:pareto_adaptation_overreliers}
     \end{subfigure}
     \begin{subfigure}{0.49\textwidth}
         \centering
         \includegraphics[width=\textwidth]{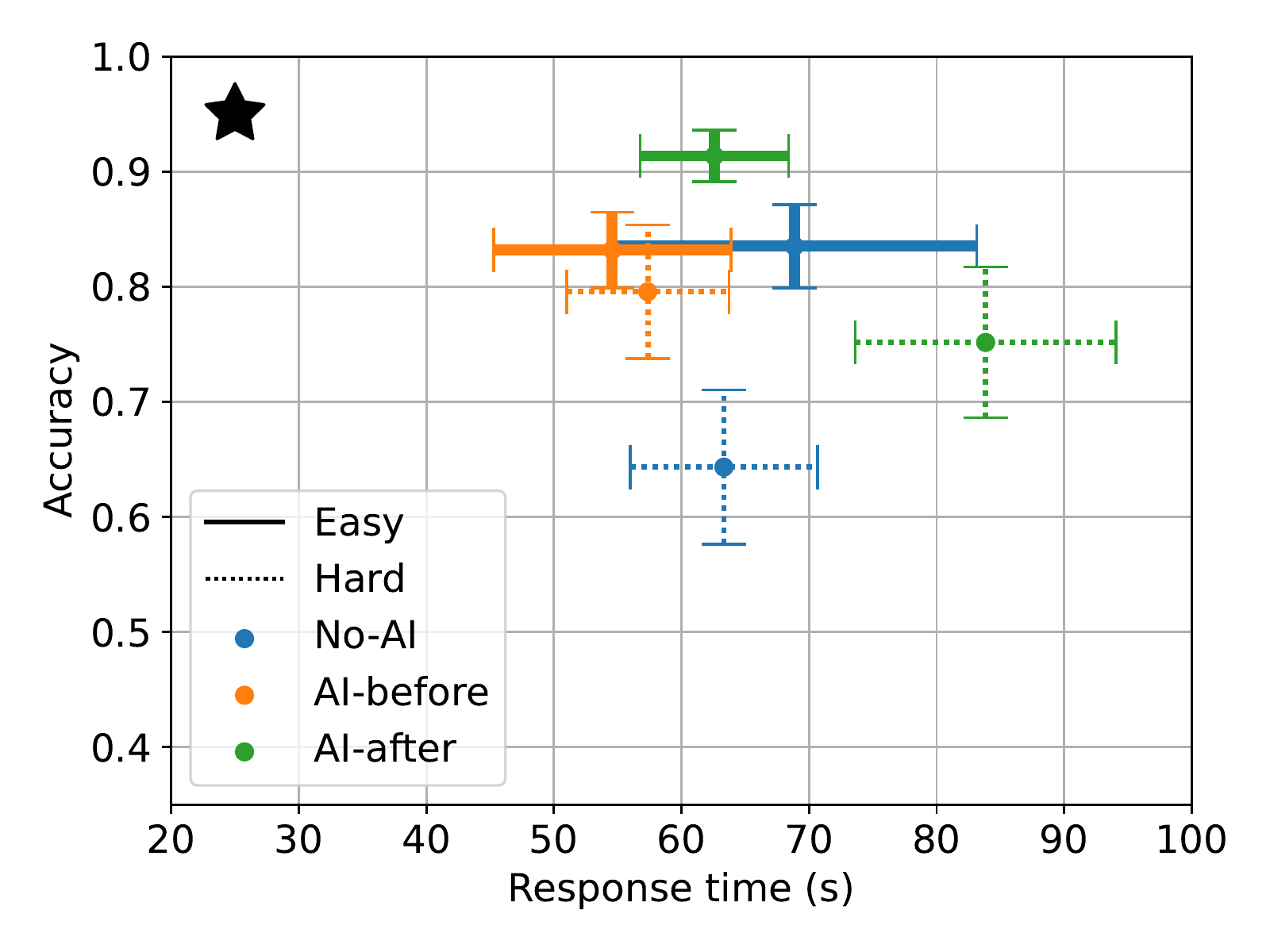}
         \caption{Not-overreliers.}
         \label{fig:pareto_adaptation_notoverreliers}
     \end{subfigure}
     \caption{
     Pareto plots: top left is desired (higher accuracy, lower response time per question). 
     We compare the performance (mean and standard error) of the AI assistances in the mixed condition on easy questions (bold lines) to hard questions (dotted lines) under time pressure.
     We see that for overreliers (left) on hard questions, there is a tradeoff between AI-before and AI-after, and for not-overreliers (right) there is a marginal tradeoff on easy questions. 
     Otherwise, AI-before is better (equal accuracy and quicker time compared to AI-after and No-AI).  
     }
     \label{fig:pareto_adaptation}
\end{figure*}

In this section, we explore how we can adapt to both the person (whether they are an overrelier or not) and the question (whether the question is easy or hard). 
We focus on the setting where people are under time pressure. 
Because we are interested in changing the assistance type depending on the question, we only look at how AI assistance types inside the mixed condition compare against each other. 
We find that we can increase accuracy of overreliers by slowing them down with AI-after on hard questions (compared to their accuracy and response time on AI-before). 
However, the not-overreliers have similar accuracy with both AI-before and AI-after, and are quicker with AI-before. 

\cref{fig:pareto_adaptation} shows the tradeoffs. 
For overreliers (\cref{fig:pareto_adaptation_overreliers}), we see that on easy questions, they are quicker with AI-before than AI-after, with similar accuracy. 
However, on hard questions, AI-after can increase their accuracy (from 55\% to 70\%) by slowing them down. 
For not-overreliers (\cref{fig:pareto_adaptation_notoverreliers}), on easy questions, they have similar accuracy and time for AI-before and AI-after (AI-after has marginally higher accuracy). 
On hard questions, AI-before helps the not-overreliers to both speed up and achieve good accuracy. 

These findings suggest that we can adapt what AI assistance we show depending on the person (overrelier or not-overrelier, which we can predict from the first part of the study) and the question (easy or hard, which we may already know, or which we can estimate, for example based on how uncertain the AI model is on the task). 
Future work can continue to look at adapting AI assistance in similar ways, and see if such adaptation improves the overall accuracy-time tradeoff. 
Future work could also look at more advanced machine learning techniques to quickly adapt, such as reinforcement learning algorithms. 

\section{Discussion}

This work focuses on the accuracy-time tradeoff of different AI assistances, and how these change under time pressure. 
Previous work compared different AI assistances focusing only on accuracy, but we consider response time to also be an important metric: for example when using AI assistance in time-pressured scenarios, or scenarios where we hope the AI assistance will speed up human decision-making. 
We specifically consider the setting where people are under time pressure (such as making decisions in an emergency room~\citep{patel2008translational,franklin2011opportunistic,rundo2020recent}), where the accuracy-time tradeoff is particularly important. 
Some recent works have shown benefits of adapting the AI assistance to maximise accuracy~\citep{noti2022learning,ma2023who,bhatt2023learning}, and we are interested in doing the same for the accuracy-time tradeoff.

\subsection{The effect of time pressure in AI-assisted decision-making}

We find that introducing time pressure can change the accuracy-time tradeoff between different AI assistances, making our AI-before condition significantly faster than the AI-after and mixed conditions. 
In our first experiment (\cref{sec:experiment1}), the same participant was alternately put under time pressure and no time pressure. 
We found that once a participant had been under time pressure once, their performance remained similar even if they were not under time pressure later in the study. 
Further work is required to investigate why the same person's behaviour changes after they experience time pressure on the task: they appear to learn how to perform the task quickly, and retain this behaviour, even if that reduces accuracy. 
In our second experiment (\cref{sec:experiment2}), we found that different AI assistances are more effective under time pressure compared to no time pressure. 

\citet{cao2023time} looked at the effect of time pressure on AI-after only, and how time pressure affects reliance on AI recommendations. 
They found that reduced observation time (before participants make an initial decision) did not consistently lead to increased reliance, while reduced decision time (the time given to a participant to update their initial decision based on an AI input) usually led to reduced reliance. 
In our studies, we did not independently test the effect of time pressure on these two phases of AI-after decision-making.  
We found that overreliance on AI-after increased marginally under time pressure, but this is likely impacted by exactly how time-pressured participants were in each phase of decision-making. 

Although we investigated the effect of time pressure in our studies, we expect that other types of pressure will also affect how people use AI assistances, such as social pressure~\citep{byrne2015chokes} or other stressors. 
Given that such AI assistance tools are increasingly being used in the workplace, where people are under different types of pressure, it is important to understand how effective different AI assistances are in such contexts.

\subsection{Overreliance rate as an important individual trait}

Previous works have found evidence that, in medical settings, experts with less domain experience trust AI more than experts with more experience~\citep{gaube2021ai,bayer2022role}. 
This may result in different overreliance rates between people. 
Although all participants are equally expert in our setting, we still find that a person's overreliance rate is stable throughout the duration of the study, and that overreliers and not-overreliers might use different AI assistances differently. 
This could be due to different factors, and future work could investigate why overreliers behave differently to not-overreliers, looking at the different mechanisms these two groups of people might be using. 
For example, are not-overreliers using AI assistance as a guide to check information, before making their own decision? 
Such mechanisms are likely to also depend on the specific task. 

Although we find evidence that overreliance is an individual trait, we find only marginal correlations between overreliance behaviour and personality traits (Big-5 Personality Traits, NFC, and self-reported belief of performance under time pressure). 
This is unlike prior work that found overreliance to be significantly correlated with NFC~\cite{bucinca2021trust}. 
This may partly be due to the nature of our task: our task was game-like and inherently motivating, meaning even people low in NFC may have been motivated to cognitively engage with the task. 
Previous work has also found that neuroticism affects people's performance on tasks under pressure~\cite{byrne2015chokes}, but we only found marginal correlations to support this. 
A potential reason is that the our personality traits were measured using only a few questions, meaning they are likely noisy estimates, and so we cannot be confident that correlations between personality traits and performance do not exist in practice. 

Overall, we believe that to successfully adapt assistance to decision-makers, we need to predict how that decision-maker will use the assistance. 
This can be done before any interaction with the system, or it can be done in real-time. 
We sought to understand if the overreliance trait could be captured beforehand (e.g., by personality variables), and were unable to do so conclusively. 
However, support can be tailored as the person interacts with AI in real-time, allowing a system to update its usage of AI assistances accordingly.

\subsection{Adapting AI assistance under time pressure}

Our findings suggest that we can adapt what AI assistance to show depending on the characteristics of the person (whether they are an overrelier or not-overrelier) and the task (easy or hard). 
For example, we can increase the accuracy of overreliers by slowing them down with AI-after assistance on hard questions, while showing AI-after to not-overreliers does not significantly increase accuracy (and only slows them down). 

Adapting AI assistance is especially important in time-pressured scenarios because achieving appropriate reliance on AI is crucial: we want to encourage a person to follow a correct AI suggestion in order to save time; conversely, when the task is more difficult and the AI assistant may be incorrect, we may want to slow down the person to ensure they still make the correct decision. 
Appropriate reliance is also important to achieve human-AI complementarity~\citep{bussone2015role, lai2019human, jacobs2021machine}. 

Previous work has argued that overreliance (or reliance) on AI can vary between different AI assistances and conditions: 
AI assistances that force people to slow down and cognitively engage with the AI assistance (like AI-after) can reduce overreliance \citep{bucinca2021trust}, and when an AI assistance is shown less often (like when adaptively showing AI assistances), people rely on it more (the scarcity effect) \citep{noti2022learning}. 
In our study, we observe the scarcity effect when there is no time pressure, but find that the scarcity effect disappears under time pressure. 
Participants already overrely more on the AI when under time pressure, and the scarcity effect does not compound with this increased overreliance. 
In our setting, participants experienced fairly strong time pressure: it is possible that, under less time pressure, there may still be a (smaller) scarcity effect. 
We also note that we found a scarcity effect with AI-before only, and did not find it with AI-after (which \citet{noti2022learning} did find). We believe this is because people needed much longer for each question in our setup (around 50-60 seconds as opposed to 10 seconds per question). 
This means that people may trust their own initial answer more in our setting, leading to less reliance with AI-after. 

We can likely further increase the performance of the human-AI team by adapting to more individual traits and properties of the task, although we did not consider them in this paper. 
First, it is possible that some people are more strongly affected by anchoring biases towards information they see, making them more susceptible to relying on AI-before, and hence changing what AI assistance is best for them. We may be able to estimate such biases using either personality traits or estimating it online, as the person interacts with AI. 
Second, a person's overall skill or knowledge of a task may affect both how much they require AI assistance, and how they use the AI assistance. 
Third, different properties of the task may matter (on top of difficulty of task): it may be more important to get certain tasks right (for example, if a patient arrives with a life-threatening condition); or maybe the AI model is known to be worse in certain settings (such as if the AI model is biased on certain tasks only), and so cannot be relied on as much. 
Adapting to all such properties is likely difficult, and requires further study and perhaps more advanced personalisation algorithms (such as using reinforcement learning algorithms).

\subsection{Limitations}

In our experiments, participants had to complete a series of logic puzzles, where all information was shown on the screen. 
This allowed us to precisely manipulate the difficulty of the task (such as by having easy and hard questions), the optimality of the AI assistance, and the form of time pressure. 
However, a key limitation of our work is that such a setting may not be realistic: in many real settings, prior knowledge of aspects of the task is important. 
Because the logic puzzles are new and different, participants may always be in System II thinking~\cite{kahneman2011thinking}, reducing the effect of cognitive forcing functions like AI-after (previous work has argued AI-after reduces overreliance because it can push people from System I to System II thinking \citep{bucinca2021trust}). 
Additionally, our task is in a non-critical setting, unlike other realistic time-pressured scenarios like doctors in emergency rooms~\citep{patel2008translational,franklin2011opportunistic,rundo2020recent}. 
Previous work has suggested that experts may have similar behaviour as participants in crowd-based studies~\citep{gaube2021ai}, but we do not know if this will also hold under time pressure. 

The AI explanation that we showed participants (along with the AI recommendation) in our setting was an intermediate symptom that led to the AI recommendation, and was easily verifiable. 
This verifiability might impact how people use AI assistances~\citep{fok2023search}.
Different forms of AI explanations may lead to different results. 
We also only compared two forms of AI assistance (AI-before and AI-after). We chose these because AI-before is widely-used, and AI-after can slow people down and potentially reduce overreliance. 
Future work could consider how people act under time pressure with other AI assistance types. 
For example, does providing only an AI recommendation (without an explanation) lead to similar results? Do other cognitive-forcing functions like ``on-demand''~\citep{bucinca2021trust} (where participants have the option to click to see the AI input) lead to similar accuracy-time tradeoffs as AI-after under time pressure?

\section{Conclusion}

Using AI assistances to help decision making in increasingly different environments and situations requires understanding the effects of the different AI assistances in each setting. 
In this paper, we consider how different AI assistances impact both accuracy and response time per question, leading to accuracy-time tradeoffs between different AI assistances. 
Such tradeoffs are especially important in situations were users are time-pressured, and we consider this in more detail. 

In our experiments, we found that introducing time pressure can change the accuracy-time tradeoff between different AI assistances, making our AI-before condition significantly faster than the AI-after and mixed conditions. 
We also found that people's overreliance rates are stable over the course of the study, and that overreliers and not-overreliers may be using AI assistances differently. 

We also found that a previously documented scarcity effect \citep{noti2022learning}, which describes an increase in reliance on AI assistance when the AI assistance is shown less often, disappears when users are under time pressure. 
This is particularly important when we are considering adapting AI assistance depending on the person and task, and further indicates that such adaptations likely need to also depend on whether there is time pressure. 
Lastly, in our exploratory analysis we found initial ways in which we might adapt our AI assistance depending on both the person (their overreliance rate) and the task (easy or hard task). 

Overall, we find evidence that different scenarios (here, the effect of time pressure) may change how people use different AI assistances relative to each other, making some better to use than others. 
This suggests that we need to be careful when applying results from one setting (such as when there is no time pressure) to another setting (such as when there is time pressure). 
Additionally, we also find evidence that we can adapt AI assistance to both user and task, adding to a growing body of literature that uses adaptive AI assistance instead of a fixed AI assistance.

\begin{acks}
This material is based upon work supported by the National Science Foundation under Grant No. IIS-2107391.  Any opinions, findings, and conclusions or recommendations expressed in this material are those of the author(s) and do not necessarily reflect the views of the National Science Foundation. ZB was partially supported by an IBM PhD Fellowship.
\end{acks}
\bibliographystyle{ACM-Reference-Format}
\bibliography{references,kzg}

\appendix

\section{Appendix}
\label{app:1}

\begin{table*}[th]
    \centering
    \begin{tabular}{llcccc}
        \toprule 
        \textbf{} & \textbf{} & \multicolumn{2}{c}{\textbf{No time pressure}}  & \multicolumn{2}{c}{\textbf{Time pressure}} \\        
        \textbf{} & \textbf{Condition} & \textbf{> No-AI acc?} & \textbf{> AI acc?} & \textbf{> No-AI acc?} & \textbf{> AI acc?} \\
        \hline
        Overreliers     & & {\footnotesize $F_{3, 78}=1.73, p=0.17$}  & & {\footnotesize $F_{3, 90}=12.4, p<.0001$} & \\
        (according to first     & AI-before & \textit{n.s.} & \textit{n.s.} & Yes ($p<0.0001$) & \textit{n.s.}  \\
        half of study)  & AI-after  & \textit{n.s.} & \textit{n.s.} & Yes ($p<0.0001$) & \textit{n.s.} \\
                        & Mixed     & \textit{n.s.} & \textit{n.s.} & Yes ($p<0.0001$) & \textit{n.s.} \\
        \hline
        Not-overreliers     & & {\footnotesize $F_{3, 82}=7.47, p=.0002$}  & & {\footnotesize $F_{3, 88}=28.1, p<.0001$} & \\
        (according to first & AI-before & \textit{n.s.} & Yes ($p=0.0063$) & Yes ($p<0.0001$) & Yes ($p<0.0001$)  \\
        half of study)  & AI-after  & Yes ($p<0.0001$) & Yes ($p<0.0001$) & Yes ($p<0.0001$) & Yes ($p=0.0068$)  \\
                        & Mixed     & Yes ($p=0.047$) & Yes ($p=0.0017$) & Yes ($p<0.0001$) & Yes ($p=0.014$)  \\
        \bottomrule
    \end{tabular}
        \caption{Comparing accuracy of the AI assistance conditions with No-AI accuracy and AI-only accuracy after splitting participants by overreliance. 
        We calculate whether a participant is an overrelier or not based on performance in the first half of the study, and calculate all accuracies only in the second half of the study. 
        We see that, both under no time pressure and under time pressure, overreliers do not significantly achieve human-AI complementarity (accuracy is not significantly higher than both No-AI accuracy and AI-only accuracy), while not-overreliers do achieve complementarity (except that AI-before under no time pressure now only has marginally higher accuracy than No-AI ($p=0.06$)).
        This is similar to \cref{table:complementarity}, but now we use the first half of the study to predict in the second half, indicating that overreliance behaviour is stable during the course of the study.}
    \label{table:complementarity_split_half}
\end{table*}

\begin{figure*}[thb]
    \centering
    \includegraphics[width=0.9\textwidth]{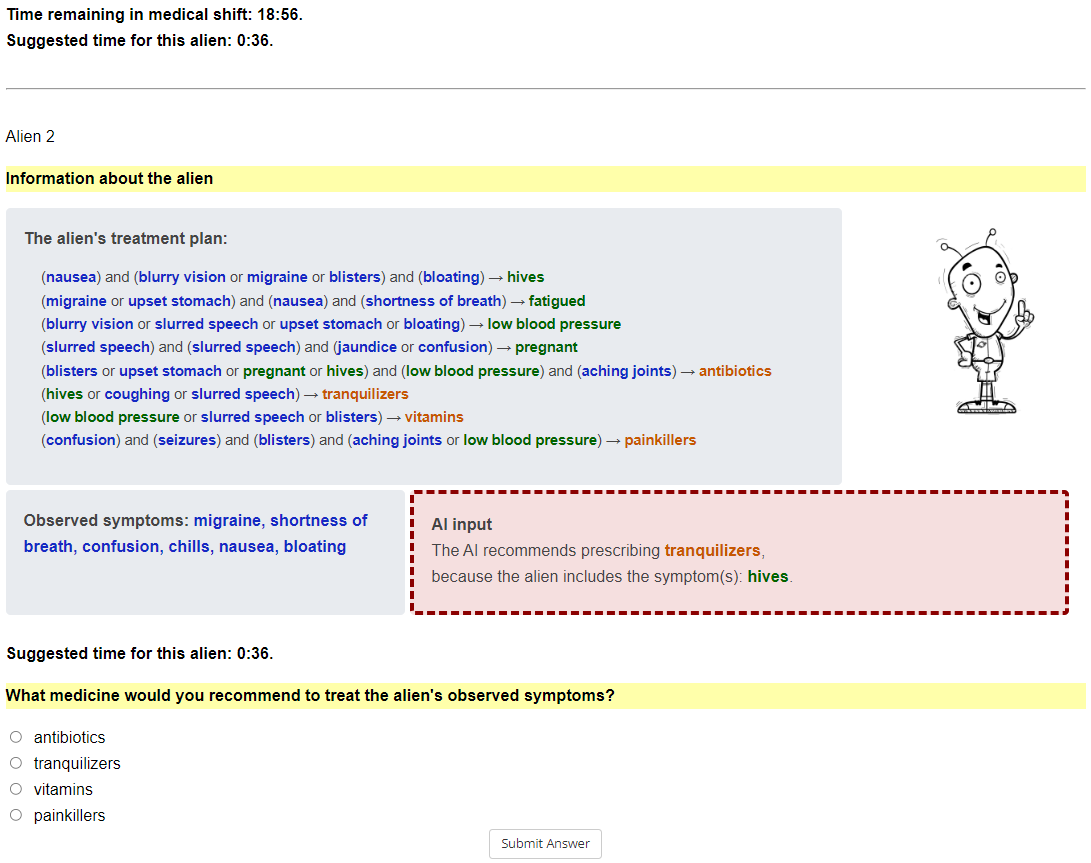}
    \caption{The alien prescription task, where participants must prescribe a single medicine. 
    In this example, the timers are shown on the screen (the global timer counts down from 20 minutes, as it is part of Experiment 2 (\cref{sec:experiment2})). 
    This question is a \textit{hard}-difficulty question, whereas the question in \cref{fig:alien_example} was easy-difficulty. 
    It is hard because the optimal medicine (which, in this case, the AI recommends correctly to be `tranquilizers') uses fewer observed symptoms than the other medicines. 
    Therefore, if a participant wants to confirm that `tranquilizers' is the optimal medicine, they have to check many other medicines too (`optimal' is defined as the medicine that uses the most observed symptoms while not using/treating any unobserved symptoms). 
    For this alien, vitamins is also a correct medicine, but it is suboptimal. 
    All other medicines are incorrect. 
    }
    \label{fig:alien_example_hard_timer}
\end{figure*}

\end{document}